\documentclass[12pt,a4paper]{article}

\usepackage{graphicx}\usepackage{amsmath}

\usepackage{subfigure}

\begin{document}

\begin{center}{\bf \Large Around the gap between sociophysics and sociology}\\[5mm]

{\large  Krzysztof Ku{\l}akowski\\[3mm]

\em {Faculty of Physics and Applied Computer Science, AGH University of Science and Technology, al. Mickiewicza 30, 30-059 Krak\'ow, Poland\\

E-mail: kulakowski@novell.ftj.agh.edu.pl\\\today}}

\end{center}

\begin{abstract}
Some basic sociophysical notions are described by a physicist, tentatively for a sociologically-oriented reader.
\end{abstract}

\section{Introduction}

It is my great pleasure to respond the kind invitation of Dr. Johannes J. Schneider to write a chapter in a book on sociophysics. Indeed, reading papers where this term is used I have often an impression that it is elusive. Now if I am, say an analytically oriented specialist in the humanities, I should devote some pages to define the sociophysics. I prefer to leave the term undefined; its picture depends on the point of view, from physics or from sociology, then here again every definition is dangerous. 

Actually, sociologists seem not to be terribly interested in the sociophysics itself; still they notice the activity of the physicists on social networks. Sometimes this activity arouses harsh comments, with reinventing wheels as rather mild. The graph on the mutual citations of the physicists and, say, sociologists working on the small world problem, shown in Ref. \cite{lcf}, shows that these groups largely ignore each other. Calls to read papers written by sociologists can be heard also from the physicists' side \cite{stff1}. In recent years a casual and loose referring to trendy sources on 'social science' (as \cite{ho1} to \cite{cowan}) was replaced by more careful quotation of basic textbooks in mathematical sociology (as \cite{ho2} to \cite{wass}). We should note that one of the leading authors in sociophysics, Serge Galam, published his early works in the Journal of Mathematical Sociology and the European Journal of Social Psychology \cite{ga1,ga2,ga4}. Also, he paid much attention to references to sociological and psychological literature. To give an example, in Ref. \cite{ga2} there are only five references to physical papers, including the original work of Ernst Ising from 1925; the other 29 references are from social sciences, mostly the social psychology. 

If some agreement is possible between the sociologists and the physicists, it is probably about the need of a more intense exchange of information. What I try here is to explain how we physicists imagine  and describe  a social system; to list standard tools of statistical physics used in sociophysical papers. I will do it frankly; if a caricature appears at the output, let it be at least clear. In this tentative description of sociology in physics (and not the opposite) I am not going to provide a complete list of references. Rather, my aim is to list most basic concepts as taken from the statistical mechanics to describe social systems: variables, structure, interaction, temperature, dynamics and phase transition. A subtitle of this text could be 'What you have to know to reject the sociophysics'. Or to accept it - as you like.

\section{Variables: spins, opinions, links and relations}

In social sciences the term opinion is yet not precise enough. In sociophysics we should then precise that we are interested in a social belief, not as commonly accepted as the property rights, but more widespread than a timetable of cat feedings. The issue should be well known but debatable, as political elections, the legality of abortion or the organization of health service. A careful analysis of this term and the like can be found in \cite{dijk}. 

In the sociophysics we represent opinions by spins. Spin is just a variable assigned to an actor or agent. There is usually a large set of agents $i$, distributed in lattice sites or graph nodes; then, we have spin variable $s_i$. As a rule in the sociophysical papers, spins can be of two values, +1 or -1, if not stated otherwise. This is what is called: Ising spins \cite{sta0}. With an increased interest of physicists in problems from social sciences the term spin was replaced by opinion or the like, but the way how it is treated in mathematical formulas and computer programs is more than often the same as in the theory of magnetism \cite{ga1,ga2, ga3,sta9,mac1}. A brief explanation of the Ising model and of  its maybe most common version - the so-called mean field theory  can be found in Ref. \cite{bsgcp}. In general, models of the ferro-paramagnetic phase transition contributed much to the sociophysical models. Basic review of the statistical mechanics of the opinion dynamics can be found in Ref. \cite{cast}; the authors write about the Ising paradigm.  

The spin variable has been generalized to capture the situations when we have more than two opinions. The spectrum of possibilities is either discrete or continuous. In both cases we can also look for magnetic analogies; the respective models are the Potts model  \cite{baxt} and the classical Heisenberg model \cite{stan} reduced to one variable. Multidimensional approach was used in \cite{kasia,sylvie} and presumably in other papers. 

Usually we assign variables to cells of a lattice or nodes of a network. Alternatively, we can assign variables to the network links. Often, we interpret such variables as the relations between the nodes. They can be large or small (strong or weak ties), positive or negative (friendly or hostile relations). In most papers the relations are discrete (plus or minus one), but some authors use also the continuous representation, as for example for the Heider balance problem \cite{mua}. The latter case is termed weighted network or weighted graph in the graph theory. In more general formulations separate sets of variables are assigned to links and to nodes. This is in accordance with the sociological textbooks on social networks, where the variables are termed structural variables and composition variables \cite{wass}.

\section{Structure} 

In sociology, the social structure is an analytical tool \cite{oxf}; this euphemism indicates that the status of the term social structure is under debate. In a geometrical sense, the structure is a set of units, with links between some of them.  References to some intermediate formulations can be found in Ref. \cite{oxf}; according to Peter Blau the structure is an ordered location of parts of the system \cite{blstr}.  As noted above, in the social network theory separate variables are assigned to the units and to the relations between them \cite{wass}. This is almost equivalent to the mathematical definition of an unweighted graph. How to assign this kind of structure to a set of social units (individuals or groups) is the matter of research in sociology \cite{wass,scot} and in sociophysics \cite{cast}. 

Recent review on complex networks \cite{bocca} written by physicists devotes five pages to the social networks; the authors refer to sociological and physical works, including the paper on the history of network analysis by Linton C. Freeman \cite{fre0}. The social networks is maybe the area of sociology where the methodology is most established \cite{wass,scot,bok}; the refereed specialistic journal Social Networks is about to enter fourth decade of its activity. There is also an International Network for Social Network Analysis \cite{insna}, and their 27-th (8-th European) conference in 2007 provided an interdisciplinary venue for social scientists, mathematicians, computer scientists, ethnologists, and others to present current work in the area of social networks. A difference in scale of effort devoted to social network analysis in sociology and sociophysics is more than obvious. 

The physicists and the like went {\it en masse} to the area of social networks mostly after the paper of Watts and Strogatz in 1998 \cite{wttstr,doro}. Duncan J. Watts is professor of sociology at the Columbia University, but he got his PhD title at theoretical and applied mechanics. In 2004 he published a paper entitled A new science of networks \cite{dwnew}, where he politely indicates the role of the physicists in this part of sociology. However, as it is clearly demonstrated in his text, it is the sociology where the basic research on the social networks was started as early as around 1951 by Rapoport \cite{rapo}. Paradoxically, four oldest papers of Rapoport referred to by Watts were published in the Bulletin of Mathematical Biophysics.

As it was noted by Watts, the (socio)physicists are fascinated mainly in simple models and interdisciplinary applications. The scale-free graphs, where the probability distribution of the degree, i.e. the number $k$ of links of a node is the power function $k^{-\gamma}$, met in their minds with the universality principle, a hypothesis known for a long time in physics of magnetic phase transitions. According to this principle the exponents (like $\gamma$ above) do depend only on the space dimensionality and the dimensionality of the magnetization. Having an information on these dimensionalities, one should be able to obtain the whole set of the exponents. Another universality in networks, what a beautiful subject of research! In this way, the networks -- social or not -- entered the computers of the physicists. Watts concludes, that 'physicists may be marvelous technicians, but they are mediocre sociologists'. For sure, in statistical physics we prefer to know something about everything than the opposite.

Are the real social networks scale-free? In small societies, the statistics is too poor to state it; in large societies it is not possible to investigate all social bonds. A notable possibility is offered by databases on phone calls and e-mail contacts \cite{hung}, but in these cases the meaning of the term contact' is somewhat twisted. There is also an obvious limitation of the number of contacts a person can maintain; this is called the Dunbar number \cite{dunb}, and is evaluated to be about 150. Therefore our question about the scale-free social networks can deal only with not-too-large societies.  

Michael Schnegg from Institute of Social Anthropology in Cologne investigated six small societies, mostly African \cite{schngg}. As the size of the societies was between 41 and 142, the comparison of the measured degree distributions $P(k)$ with the power law function $k^{-\gamma}$ was rather provisional: some plots were maybe similar, some other  not at all. However, this evidence allowed to find a correlation between the character of the plot and a purely sociological quantity: the tendency to give to those from whom one received in the past. Research on this tendency the so-called reciprocity  has a long tradition in social sciences \cite{mali,mauss}. In this way, the description of the mathematical structure of the investigated societies got a truly sociological dimension.

\section{Interaction} 

In physics, the only way for an object to exist is to have energy: a possibility of doing work. In social sciences, there is nothing like energy: no energy conservation, no energy measurement. In physics, the way how to detect the interaction between objects is to observe correlations between their states. In this sense, the physical interaction is detectable also in sociology.

According to Piotr Sztompka, in sociology there are four levels of theoretical description of interpersonal interactions \cite{szt}. At the first level the interaction is described as a chain of impulses and reactions. This theory benefits much on the behavioral psychology and experiments with rats \cite{lndz}. At the second level the theory of social exchange, usually associated to George C. Homans \cite{homa}, bases on the assumption of rational choice and uses the formalism of the game theory \cite{neuma,straf}. The above mentioned concept of reciprocity applies right here. At the third level, meanings of interactions are of concern as important determinants of human contacts. The partners currently analyse the content of the information which is mutually sent. If the meanings of signals match, the contact can develop. The theory is termed symbolic interactionism and it was created by Charles H. Cooley and George H. Mead. At the last level of abstraction we find the theory of social roles and labeling by Erving M. Goffman. The present author feels that it is appropriate to abstain from describing this theory in one sentence, it is better to refer to the literature \cite{goff}.

In the sociophysics, most of these subtleties are lost. Within the magnetic analogy, the interaction between partners is reduced to the energy of a mutual interaction. Once the states (opinions) of the partners 1 and 2 are represented by their spins $s_{1,2}=\pm1$, the interaction can be represented by the Ising term $-Js_1s_2$. For $J>0$ this interaction favors the same opinions of the partners; for $J<0$ - the opposite opinions. More examples of the applications of the Ising model can be found in \cite{sta0}. Results slightly more general can be obtained with using a utility function \cite{cole} which plays the role of negative energy; the action of the partners can be reduced to attempts to get a maximum of  this function.

If one intends to apply the game theory, the concept of the utility function fails except for a narrow family of games, the so-called potential games \cite{szabo}. An example of the potential games is the so-called network congestion game; there, agents look for the shortest path to reach resources, distributed on a network \cite{ncggm}. The interaction appears when two agents use the same path, which is then more costful. For a review of games played on networks we refer to Ref. \cite{szabo}. Among other games, the famous Prisoner's Dilemma is discussed there. It is known to provide frames for numerous problems in ecology, international disarmament, marketing and other areas \cite{straf}. 

The number of methods how to model the mutual influence of partners of a social interaction in computers is potentially unlimited. In the voter model, an agent picks at random the opinion of a neighbor \cite{cliff,cast}.  In 1981 the psychologist Bibb Latan\'e (now Director of his own Center for Human Science) published an influential paper with mathematical formulation of the theory of social impact \cite{bibb}. Later his findings were developed, formalized and simulated on computers \cite{nwk1,nwk2}. As variables, the authors use attitudes $\pm 1$, and the changes of these attitudes depend on the persuasive and supportive impact parameters of the partners. It is not far from these formulations to more recent ones. In the Sznajd model \cite{sznjd}, accordance of opinion of a pair of neighbors causes that this opinion is shared also by other neighbors of the pair. In the Deffuant model \cite{deff} opinions are represented by real numbers. If opinions of two neighbors happen to be not too different, they approach each other even closer. In the Hegselmann-Krause model the same yes-if-not-too-far rule applies to all neighbors simultaneously \cite{KH}. In the Galam model agents meet in small random groups and adopt the opinion of the majority \cite{galmaj}. For a review we refer to \cite{cast}, noting that the responsibility of the sociophysicists for all that jazz is at least shared with professional specialists in the humanities.

To summarize this section, in the sociological theory of networks the interaction strength between two actors' network nodes -  is expressed by the value of a variable assigned to the link  or the tie - between these nodes \cite{wass}. According to the famous sociologist Mark Granovetter,  this single variable should express a (probably linear) combination of the amount of time, the emotional intensity, the intimacy (mutual confiding), and the reciprocal services which characterize the tie \cite{gran1}. Granovetter demonstrated, that new informations in a social network arrive usually by links which join otherwise unconnected parts of  the network. Such links are usually termed weak; the role of weak ties became a cornerstone in the social network theory. 

\section{Temperature, information and noise} 

In physics, temperature is inseparably connected with energy. In thermal equilibrium the probability of appearance of a given state depends on the ratio $E/T$, where $E$ is the energy of this state. As noted above, there is no energy in social sciences; then it seems that the temperature cannot be present there. However, there is also a more general view on the temperature, which connects it with information or its lack \cite{mach}. This aspect can be useful here.

Suppose that the information is available about the actual state of a given social issue. To give an example, everybody knows that the police efficiently controls the city. This means that the uncertainty about the possible action of the police is negligible. Bill wonders what if he robs the bank at the city center. He finds that with probability close to one, practically for sure, he will go to jail. The temperature assigned to this situation is zero; $T=0$. However, if nothing can be told about his future, then $T$ is very large; all states are equally probable. At intermediate situations, $T$ is also intermediate. 

Having this aspect of temperature in mind, we can accept that some energies are assigned to tentatively social states. High energy of a state means only that this state is considered as improbable. On the contrary, low energy means that the state will appear rather frequently. As a rule, the state with the lowest energy (the so-called ground state) appears with the highest probability. Then, temperature is an important measure of the probability ratios. For $T=0$, it is only the ground state which can be observed. For $T$ infinite, nothing is known. As it is known in social sciences, things perceived as real are real in their consequences \cite{thth}. It is then of secondary importance, if there is somebody in the society who knows more and who can assign smaller temperature to the relevant probability distribution. Actors evaluate the probabilities according to their actual knowledge. In this sense different temperatures can be assigned to different actors. Their states of being informed or being uninformed is a social fact. Temperature can be then used to describe a willingness to behave randomly, against rules. This trick has been used to improve the original Schelling model of the urban segregation \cite{schel,staso}.

At equilibrium, temperature in the system is equal to the environment temperature, this is an ultimate fate of every finite physical system. (A question remains, how long it takes to make it equal?) This leveling can be also treated as a rule when we deal with information, if we believe that every secret will be known after long time. Before it happens, various temperatures can be assigned to various actors. This option was used in Ref. \cite{mull}, where inhomogeneities of $T$ were shown to lead to some kind of self-organisation.

The self-organized criticality \cite{soc} is of interest for itself in statistical physics. In this state, the probability distribution of random fluctuations is a scale-free function of the fluctuation size. This means, that the fluctuation size is not limited from above. In particular, changes of a social system in this state can be arbitrarily large; indeed there is an experimental evidence that in some conditions a society is at the border of a limitless catastrophe \cite{iraq}. Recently, the dynamical behavior of a standard Ising system in these conditions was simulated \cite{staku}; there, the temperature was a measure of intensity of the self-organized noise. In general, by temperature one can mean an amount of uncertainty. This is an analogue to some physical probability distribution functions, where in some cases the width of a maximum of the function increases with temperature.

\section{Phase transition, an equilibrium state and the thermodynamic limit} 

What a sociophysicist likes most to obtain is a phase transition. The reason is that he knows that social theories are qualitative. According to a common knowledge, the phase is 'any of the forms or states (...) in which matter can exist' \cite{common}. A change of phase, as for example a revolution or a paradigm shift, should  be perceptible in a society even without measurements. That is why phase transitions are discussed so often in sociophysical papers \cite{ga7,getto}. 

However, theories of phase transitions demand some ingredients which are impossible to occur in the social world. First is the so-called thermodynamic limit; the transition is well defined only in infinite systems. In an infinite system there is always a finite probability that the phase will be changed due to fluctuations of some uncontrollable quantities. This is particularly true in social reality, where each system has its finite lifetime; from months for a couple of lovers to centuries for an empire. Recently, a sociophysical model was designed of a network of cooperating scientists or laboratories \cite{hoho}. There, several phases of the network were distinguished, and the system was able to change its phase in some kind of a random walk in the phase space. In my opinion, this approach reflects the unpredictability of social systems.

It is also possible that some kind of change of state appears in a finite system, but the transition point (as the critical level of noise) vanishes or tends to infinity when the system size increases \cite{fs1,fs2}. These cases cannot be termed phase transitions within the standard theory. However, the effects can be meaningful in social sciences.  

Second theoretical ingredient, connected with the previous one, is the condition of equilibrium. In physics, the notion of equilibrium, although only intuitive, is well established. In a social system, equilibrium is never attained. Some correspondence to this difficulty exists also in the statistical physics, for example in the theory of spin glasses. There, various processes proceed with different velocities, from very quick to so slow that the laboratory went bankrupt before the equilibrium is attained.  In such cases one usually intends to assume that there are two kinds of processes: very quick  and those are over before an experiment has started, and very slow  and those do not change the system state while data are gathered. It is not always the case that this assumption works, but maybe it is unavoidable in social experiments.

\section{Dynamics}

Statistical mechanics of the non-equilibrium processes is much less developed than the equilibrium theory  and this is not strange, because in the former case the aim is much more ambitious. It is far from clear to what extent deterministic dynamic processes can be described with any kind of averaging over the state space. Basically, we rely on the theory of random walk, powerful and rich in applications \cite{kamp,redner}. Two particular methods of this theory are applied in sociophysics relatively often: the master equations \cite{weidlich,helbing} and the Monte-Carlo simulation \cite{stauffer}. First method is analytical, relatively easy and approximate, second is numerical, technically difficult and in principle exact. 

Numerous social processes were attempted to be described with these methods: migration dynamics \cite{weidlich}, residential segregation \cite{stff1}, competitions \cite{bona}, gossip \cite{lind}, evolution of cultures \cite{cult} and languages \cite{lang}, opinion dynamics \cite{opidyn,opiger}, innovations \cite{jain} and many others. All that is not far from the research done by non-physicists: \cite{hm1,hm2,axe,hmgoss,hmcult,hmlang,bibb,hmopidyn}. The flow of ideas is multidirectional: references to complexity, diffusion, entropy, self-organization, fluctuations, fractals, randomness, criticality and chaos can be found in papers on economy and social sciences as well as in lists of physical key-words. 

In sociological textbooks \cite{szt,tur,castells} motifs of change, evolution, dynamics and progress are basic.  
Social theories of change grapple with the separation of what is constant from what is changing, what is cyclic from what is unique \cite{gies}. One can imagine that we could be faced with similar problems in physics if we decide to investigate a medium of some tens of objects, interacting but well identified, distinguishable and with some memory. It is not clear if the dogma of free will could make these problems seriously worse. Personally, I would rather suspect that this dogma could be easily removed by some smart approximation.

\section{Final remarks}

In sociophysics, physical notions and models are applied to social phenomena. This means also that the sociophysicists treat the society as a physical system. We look for a diffusion of opinions, for a hysteresis of strikes and for phase transitions in cultures as if they appeared in magnets. Even human bodies can be treated as a physical material, with appropriate friction coefficients, to describe the behaviour of crowds in cramped space \cite{vic}. It is not surprising that in this research, experiments work better if light is switched off. In general, a model system is less complex if agents are less endowed: less strategies, shorter memory, information limited to nearest neighbors, etc. In physics, we are trained to make our models as simple as possible. Then, often we are tempted to approximate human minds by Something Somewhat Simpler.

As noted in the Introduction, the sociophysicists are rightly criticized by the social scientists for not referring to the sociological papers. There is also another tension between the empirical and the hermeneutic sociology. In the latter, the sociophysics will be at best ignored. Piotr Sztompka - leading Polish sociologist - puts the behaviorism, Homans, Blau and Coleman all together to a bag of epigones of positivism (\cite{szt}, p.27). In a recently edited set of basic texts \cite{sztku}, he quotes the essay of Stanisaw Ossowski from 1962 about differences between natural and social sciences. The Ossowski's arguments are \cite{oss}: self-fulfilling or self-destroying prophecy, influence of the measurement on the measured properties of the system (vide: the Heisenberg uncertainty principle), the conflict between specific character of investigated societies and the need of universal results, the general lack of insight of standardized statistical methods, the projection of the internal world of the researcher on his results. To quote these arguments now means to treat them either as currently valid or as a reference point for further discussion. 

Let us explore the latter possibility. We are going to contrast at least some of the above arguments with recent text of Adam Grobler \cite{gro} on the methodology of sciences. According to his formulation, the opposition is to be built between naturalistic and anti-naturalistic point of view. To start from the former, the essential aim of science is to discover laws which are experimentally verifiable and enable us to explain and predict the observed phenomena. Further, there are three basic arguments against the naturalistic approach. These are: 1) social processes are too complex and mutually entangled, that the above given aim cannot be carried out; 2) social phenomena do depend on human will and as such, they are unpredictable; 3) in contrast to nature, people can take into account the social theory, what influences the reality described by this theory.

As we see, point 3 matches with the first objection of Ossowski. As it follows from further parts of the Grobler's text,  it can be dismissed by construction of a series of theories, $T_n$.  Each theory $T_{i+1}$ takes into account the influence of previous $T_i$ on the human behaviour. In the limit we get the true theory. Let us add that temporal theories can also be useful as approximations, at least when describing behaviour of people less up-do-date with the current state of sociological art. Similar argument is working for the second objection of Ossowski, usually termed the interviewer effect. In this case, the problem is even easier, as the measurement (polls) never touch the whole population. If the poll result is published, again a next-step-theory can take its influence into account.

The necessarily specific character of laws drawn from the observation of a specific society reminds the pre-scientific, i.e. technological state of physics \cite{tatar}; in this sense the sociology can be seen as young in methods (but not in problems discussed). However, we know that the related problem of sample-dependence of the measured properties is eternal in the solid state physics. In the statistical physics, the related phenomenon is investigated under the name lack of self-averaging, and its appearance can serve as a fingerprint of the critical states. Here, the point of view of a physicist can provide a new insight. The next problem is how to interpret the results obtained with the standardized statistical methods rather than whether to apply them; here the sociophysics is in good company. Our way is to improve the methods to be applied and not to resign beforehand.

The last problem raised in the Ossowski's text does not apply only to sociophysics or to naturalistically oriented sociology, but it threatens the whole science \cite{gro,tatar}. As such, it cannot be applied to deny the validity of naturalistic sociology, if it does not invalidate the naturalism as a whole. Therefore, the discussion of this point must be moved to another place. Coming back to points 1) and 2) in the Grobler's text, the author rightly indicates that the objection on the complexity of social systems applies also to meteorology; still nobody denies the need of weather forecasting, even if the results are doubtful. Similarly, the free will of human beings does not preclude the predictability of the human behaviour. 

This point 1) in connection with 2) deserves a broader comment. It seems that the arguments against application of natural sciences to social systems base on the view of natural sciences in positivist frames. Indeed, as it became known around 80s \cite{sag,lei}, the deterministic character of laws of physics which rule a dynamical system does not suffice to predict the system behaviour. The effect is known as deterministic chaos. This allows to continue the argument of Grobler, related to his point 2). Grobler says: laws of nature do not lead to determinism. We can add: determinism does not lead to predictability. Let us note here that in his textbook on sociology \cite{szt} Piotr Sztompka makes a remark on the so-called butterfly effect. This anecdotical effect (a move of butterfly's wings can affect the weather) is an example of the sensibility of a physical system behaviour with respect to its initial conditions; a fingerprint of the deterministic chaos. 

From my point of view, the main postulate of the hermeneutic sociology is to admit that people have minds and use them. Certainly we sociophysicists should be then more careful when justifying our simple models, where sometimes this action of minds seems to be averaged out and finally removed by virtue of the law of large numbers. Personally however I believe that there is no contradiction between the hermeneutic sociology and the statistics. The gap between the naturalistic or empirical sociology and the sociophysics is perhaps smaller than the one between the naturalistic and the hermeneutic or humanistic social science. A tolerant sociologist can treat the sociophysics as a part of the mathematical sociology. We sociophysicists should also feel to be a part of the whole, distinguished by some bias towards using ideas and formalisms from the statistical mechanics. Can this bias be fruitful? While the positive answer could be debatable, the negative answer is premature. 

Concluding, it seems possible and desired that the boundaries between sciences will be considered more as determined by methods, and not by subjects of research. As good methods flow to other areas, we could evolve to Unified Knowledge. This would be more modest than to declare a new kind of science (reference removed). In my opinion the validity of all these divisions is ultimately minor, when they are contrasted with our common curiosity and interest in social laws. It is impossible to predict the future state of our knowledge \cite{sztpop}; the same does apply to the ways to advances in science.

\end{document}